\documentclass[a4paper, 10pt]{article}
\usepackage[top=2cm, bottom=2cm, left = 2cm, right = 2cm]{geometry} 
\usepackage[utf8]{inputenc}
\usepackage{textcomp}
\usepackage{graphicx} 
\usepackage{amsmath,amssymb}  
\usepackage{bm}  
\usepackage{circuitikz}
\usepackage[pdftex,bookmarks,colorlinks,breaklinks]{hyperref}  
\usepackage{memhfixc} 
\usepackage{pdfsync}  
\usepackage{fancyhdr}
\usepackage{pgfplots}
\usepackage{gensymb}
\usepackage{pgfplots}
\usepackage{tablefootnote}
\usepackage{tikz}
\usepackage{float} 
\usepackage[numbers]{natbib}
\usetikzlibrary{positioning}
\usetikzlibrary{patterns}
\usepackage[toc,page]{appendix}
\usepackage{amsmath}
\usetikzlibrary{shapes,arrows,positioning}

\tikzset{
block/.style={
  draw, 
  fill=blue!20, 
  rectangle, 
  minimum height=3em, 
  minimum width=6em
  },
sum/.style={
  draw, 
  fill=blue!20, 
  circle, 
  },
input/.style={coordinate},
output/.style={coordinate},
pinstyle/.style={
  pin edge={to-,thin,black}
  }
} 
\title{Is it Truly Necessary for Bicycle Power Meters to Rapidly Sample Angular Velocity?}
\author{Jack Renshaw - The University of New South Wales}
\begin{document}
\maketitle
\bibliographystyle{ieeetr}
\bibdata{sources}
\begin{abstract}
Bicycle Power Meters have become ubiquitous in professional and amateur cycling. These devices claim
high levels of accuracy, and this accuracy is indeed essential if they are to serve their purpose as reliable
training aids and indicators of improvements in fitness. Power is generally obtained via the independent
estimation of torque and angular velocity. Designs vary in the way in which they estimate 
angular velocity. Some power meters estimate angular velocity many times a second, whereas
other power meters compute an average value for each pedal stroke.
The aim of this paper is to investigate whether it is necessary to rapidly sample
angular velocity in order to obtain accurate power values under conditions of dynamic equilibrium. Countering previous research on the topic, this paper finds 
that average angular velocity alone is usually sufficient for the purposes
of computing power, although there may be certain exceptional circumstances where consideration
of harmonics may appreciably improve fidelity.
\end{abstract}
\tableofcontents
\section{Introduction}
In his preface to \textit{Electromagnetic Waves},
Sergei Schelkunoff noted that
theoretical physicists have a tendency to \textit{"lose sight of the interdependence of force and velocity waves"}
when considering the propogation the eponymous waves through various media,
and that the engineering approach, which was to \textit{"introduce the second important wave concept, that of the impedance"},
proved to be of significance in physics as well as engineering.\\
When designing electromechanical devices that aim to accurately estimate power,
the product of force and velocity, it is no less important to consider, in a concrete fashion,
the mechanical characteristics of the power producing system
and characteristics of the medium through which the system moves. This paper will suggest
that an appropriate analytical framework for modelling such systems
and their environment is via an electrical analogue,
thereby leveraging the concept of impedance and the plethora of techniques
and simulation tools available to electrical engineers. Prudent consideration of
the impedance of the system of interest is necessary to
ensure that power estimating devices perform, and to ensure that they are not over-engineered.\\
This paper will focus on the esimation of mechanical power
by bicycle \textit{Power Meters}, however the framework and methodology
utilised by this paper is applicable to a number of situations
beyond this particular context.
In contrast to previous research on the topic, 
this paper concludes that, under most circumstances,
the mechanical and biomechanical characteristics of the typical cyclist are such 
that the rapid sampling of angular velocity is not necessary to obtain
highly accurate power estimates; accurate power estimates
can be obtained by considering only the average angular velocity throughout the pedal stroke. 
This result derives from the fact
that cyclists have high inertia. Even when variations in velocity or force are high,
this inertia tends to result in the principal harmonic components of force and velocity
co-existing in quadrature, and they therefore make no meaningful no contribution to mechanical work.
\subsection{Power Meters in Ameteur and Professional Cycling}
Bicycle Power Meters have become a mainstay of effective training
for professional and amateur cyclists. These devices aim to estimate
the mechanical power delivered by the rider to the bicycle and ultimately
to forward propulsion.
The purpose of estimating power is to deliver an objective measure
of performance to a sport that is confounded
by complicating environmental variables.
An ideal power meter would measure and report the power applied
to the pedal at every point. As it is not possible to measure power 
directly, power is usually estimated via the independent estimation of force and velocity.
Power meters generally function by estimating torque
via the use of strain gauges, and estimating angular velocity
either via the measurement of rotational period, or
centripetal acceleration. These quantities are related by the following equation:
\begin{equation}
  P = \tau \times \omega
\end{equation}
Power is typically calculated over the period of a single pedal stroke,
which is the period ($T$ henceforth) over which the bicycle crank traces an entire
revolution. Torque and angular velocity may be sampled spatially or temporally.
Properly computed, the average power over this period is:
\begin{equation}
  P_{avg} = \frac{1}{T} \int_{0}^{2 \pi} \tau(\theta) \times \omega(\theta) \: d \theta = \frac{1}{T} \int_{0}^{T} \tau(t) \times \omega(t) \: dt
\end{equation}
\subsection{Literature Review}
This paper is largely a response to, and critique of,
a research paper published by Favero\cite{favero}. Favero is a manufacturer 
of a popular power meter, the \textit{Assioma}.
One crucial differentiating feature of the \textit{Assioma}
in relation to other power meter models
is \textit{Instantaneous Angular Velocity} (IAV),
which is essentially a high rate of sampling angular velocity;
a feature achieved through the use of gyroscopes.
The Favero Paper seeks to justify the supremecy of this design choice, and contains two aspects:
\begin{enumerate}
  \item A theoretical discussion of how pedal stroke variation might give rise to error in the computation of power.
  \item Analysis of experimental data that supposedly demonstrates that fact.
\end{enumerate}
Favero's Paper concluded that
not considering variations in angular velocity could 
account for errors \textit{"as high as -1.6\% using round chainrings and +4.5\% using oval chainrings"} and that \textit{"different patterns were also observed depending on the type of indoor trainer used"}.\\
Danek et Al, 2022\cite{danek2020modelling} expanded upon the brief theoretical explanation provided by Favero with deeper analysis,
and then provided a numerical example which indicated that there existed 
\begin{quotation}
  \textit{a discrepancy between the power-meter calculations resulting from the use of the instantaneous-pedal-speed and average-pedal-speed information. Removing this discrepancy is crucial for a variety of information that rely on power measurements, as is the case of this article}.  
\end{quotation}
Danek et Al stated that \textit{"neglecting speed variation during a pedal stroke"} produced a \textit{"different - and less accurate"} power result.\\
This paper will introduce three primary criticisms of the Favero Paper on
\textit{Methodological}, \textit{Inferential} and \textit{Theoretical} grounds.
These critiques are based on what this paper finds to be 
the errant conclusion of the Favero Paper, which is that:
\begin{quotation}
\textit{variation of angular velocity within a rotation has a considerable impact on the error in calculation of any algorithm that does not take it into account.}
\end{quotation}
This claim implies a causal relationship between angular velocity variation and power error.
The paper further claims that:
\begin{quotation}
  \textit{all cycling conditions that produce an increase in the unevenness of rotational speed within the pedal stroke progressively influence the error in the calculation of power; these conditions depend on the equipment used (chainrings), cycling situation (on trainer or road, flat road or climb), as well as the pedaling style of each cyclist.}
\end{quotation}
The use of "progressive" implies a 'dose-response' relationship between angular velocity variation
and power error.
  The primary empirical basis for this claim is the \textit{"high correlation coefficient [of] $R^2 = 0.93$"} between 
  Angular Velocity and Power Error. The Paper also states that:
\begin{quotation}
  \textit{only signals relevant to the left pedal were aquired}
\end{quotation}
However, Favero's Paper also claims that:
\begin{quotation}
  \textit{all the following considerations [related to power error being produced by variation in angular velocity], although referring to the pedal, remain valid even when the points at which the moment of force M and the rotational speed $\omega$ are detected are situated on the crank arm or spider or on the bottom bracket/hub.}
\end{quotation}
While it is true in the strictest sense that AV Variation is necessary for power error to occur,
this paper will argue that it is neither 
causal (i.e, AV Variation is the by-product of system dynamics, and it is the system dynamics
that give rise to power error if and where it exists), nor is AV Variation alone
sufficient to produce meaningful power error.
Furthermore,
this paper will argue that power error observed by Favero's Paper
under certain scenarios was a side effect of the 
test design (specifically the choice to 
compute power unilaterally - refer to Section \ref{sec:unilateral}), and quantative conclusions in relation
to power error cannot be generalised to most power meters, which measure power bilaterally.
\subsection{The Operation of Bicycle Power Meters}
This paper will adopt Favero's Paper's functional categorisation of
techniques for measuring angular velocity for the purposes of computing power; IAV and AAV. These 
techniques are outlined below.
\subsubsection{Instantaneous Angular Velocity (IAV)}
Power Meters that rapidly sample angular velocity typically 
utilise multiple gyroscopes or accelerometers to estimate angular velocity
instantaneously, usually at the same rate at which strain is sampled.
Power under the scheme of IAV can be computed by taking the 
average value of the product of the constituent terms:
\begin{equation}
  P_{avg} = \frac{1}{N} \sum_{t=1}^{N} \tau[t] \omega[t]
\end{equation}
\subsubsection{Average Angular Velocity (AAV)}
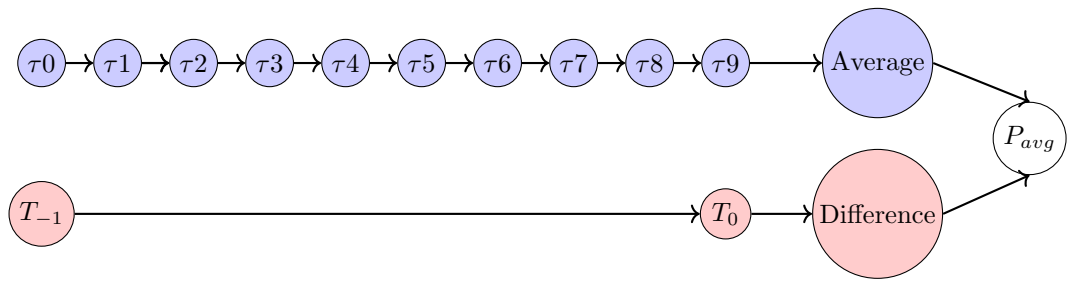
\begin{figure}[h]
  \centering
  \begin{tikzpicture}
    \foreach \i in {0,1,2,...,9} {
      \node[circle, draw, fill=blue!20, inner sep=2pt] (sample\i) at (\i, 0) {$\tau \i$};
    }
    \foreach \i in {0,1,2,...,8} {
      \draw[->, thick] (sample\i.east) -- (sample\the\numexpr\i+1\relax.west);
      }

    \draw[->, thick] (sample1.east) -- (sample2.west);

        \node[circle, draw, fill=red!20, inner sep=2pt] (sample11) at (0, -2) {$T_{-1}$};
        \node[circle, draw, fill=red!20, inner sep=2pt] (sample12) at (9, -2) {$T_{0}$};
    
        \draw[->, thick] (sample11.east) -- (sample12.west);

        \node[circle, draw, fill=blue!20, inner sep=2pt] (process1) at (11, 0) {Average};
        \node[circle, draw, fill=red!20, inner sep=2pt] (process2) at (11, -2) {Difference};
        \node[circle, draw, fill=white!20, inner sep=2pt] (process3) at (13, -1) {$P_{avg}$};

        \draw[->, thick] (sample9.east) -- (process1.west);
        \draw[->, thick] (sample12.east) -- (process2.west);

        \draw[->, thick] (process1.east) -- (process3.north);
        \draw[->, thick] (process2.east) -- (process3.south);

      \end{tikzpicture}
  \caption{System model for the measurement of power via AAV}\label{fig:aavscheme}
\end{figure}

Some power meters utilising AAV perform an averaging of sampled angular velocity over the 
course of the pedal stroke, whereas other designs only have the ability to determine angular period.
The general scheme of AAV-based power measurement is shown 
in Figure \ref{fig:aavscheme}.
Essentially, torque $\tau$ is sampled at a high rate, and a time-difference is computed
for the length of a pedal stroke. 
\\
The combination of time and torque measurements result in an "average" power figure via
homogenously weighting every sample of torque taken:
\begin{equation}
  P_{avg} = \frac{\sum_{t=0}^{N}\tau_{t}}{N*(T_{0}-T_{-1})}
\end{equation}
\subsubsection{Theoretical Case for Power Error}
As indicated by the Favero Paper, the AAV technique neglects any consideration of 
angular velocity variation throughout the pedal stroke. Specifically,
this technique assumes either no variation in angular velocity
throughout the pedal stroke, or that if there is angular velocity variation,
it is uncorrelated with the application of torque.
Favero's argument that this will produce power error is summarised as follows:\\
Power can be expressed as the cross-product of torque ($\tau$) and angular 
velocity ($\omega$) , which vary in time.
\begin{align*}
 P(t) = \tau(t) \times \omega(t)
\end{align*}
Both $\tau$ and $\omega$ \textit{"can be expressed as a Fourier series of sine waves (harmonics) extending by periodicity a single period"}\footnote{This claim makes the unstated assumption that the Fourier Series of the constructed periodic signal converges, which assumes smoothness. As will be discussed in a later section, this implies conditions of dynamic equilibrium}.
For simplicitly, $\tau$ and $\omega$ will be treated as scalar quantities by considering
only rotation about the crank spindle.
A substitution of the Fourier Series of these signals into the power equation yields:
\begin{align*}
  P(t) = \tau_{0} * \omega_{0} + \\
  \tau_{0} * \sum_{n=1}^{\infty} \omega_{n} \cos(\frac{2\pi n t}{T} + \theta_{n}) +\\
  \omega_{0} * \sum_{k=1}^{\infty} \tau_{k} \cos(\frac{2\pi k t}{T} + \phi_{k}) +\\
  \sum_{n=1}^{\infty} \sum_{k=1}^{\infty} \tau_{k} * \omega_{n} * \cos(\frac{2\pi n t}{T} + \theta_{n}) * \cos(\frac{2\pi k t}{T} + \phi_{k}) 
\end{align*} 
Power error derives from the final double-sum term\footnote{The Paper performed a trigonometric substitution to reduce the double-sum to a single-sum}, as the preceding
terms evaluate to zero when integrated over the period of the pedal stroke. The Favero Paper also notes the following:
\begin{enumerate}
  \item The contribution of harmonics (of the same order with eachother) is zero when they are in quadrature.
  \item Angular velocity tends to have a dominant harmonic twice the fundamental frequency.
  \item It is not possible to establish a priori whether the harmonics can increase or decrease the result obtained from the average values.
\end{enumerate}
The primary critique of this theoretical argument is that the relationship
between torque and angular velocity is decontextualised when circular chainrings are used. Under the conditions implied,
where the 'periodicity of a single pedal stroke is [smoothly] extended',
variations in angular velocity are \textit{induced} by variations in applied torque,
and it is in fact possible to make reasonable predictions about both the sign
and magnitude of power error. Specifically:
\begin{enumerate}
  \item The fact that angular velocity lags force (by at most quadrature), means that AAV will necessarily under-predict power. Power Error will thus be strictly negative\footnote{Adopting Favero's convention for error sign}.
  \item Cyclists tend to have high inertia, so harmonics of the same order are usually in approximate quadrature.
  \item By the fact of 2, power error under conditions thus described is generally very small.
  \item When inertia is low, for example on an indoor trainer, power error may genuinely be elevated.
\end{enumerate}
\section{Analysis of Favero Study and Dataset}
A flaw in the Favero Paper stems from the improper treatment of collected data,
the theoretical and methodological limitations noted above notwithstanding. The authors misinterpreted
the collected data by drawing a false continuity between the angular velocity
variation induced by ovalised chainrings and that induced by circular chainrings,
thereby attributing power error induced by another aspect of ovalised chainrings
(namely, a changed phase-angle relationship) to the increased variation in
angular velocity which is a side-effect of ovalised chainrings.
The Favero Paper's dataset provided in the study was entered into Python for reanalysis.
producing a near-identical result.
A strong Cubic relationship between
angular velocity variation and power error is visible. The data was also reanalysed using a Linear Fit. 
\begin{figure}[H]
  \centering
  \includegraphics*[scale=0.5]{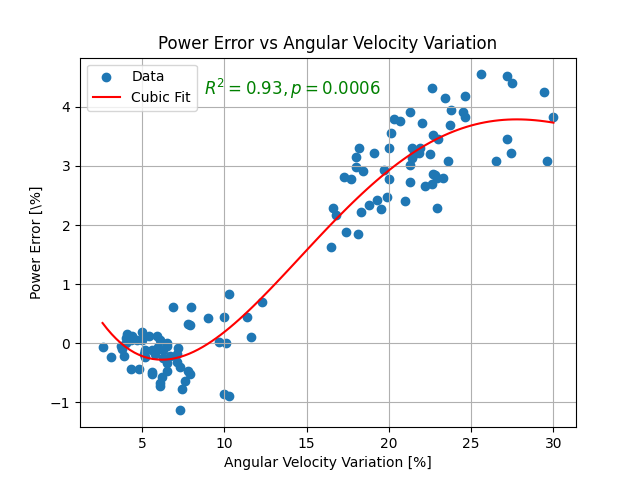}
  \includegraphics*[scale=0.5]{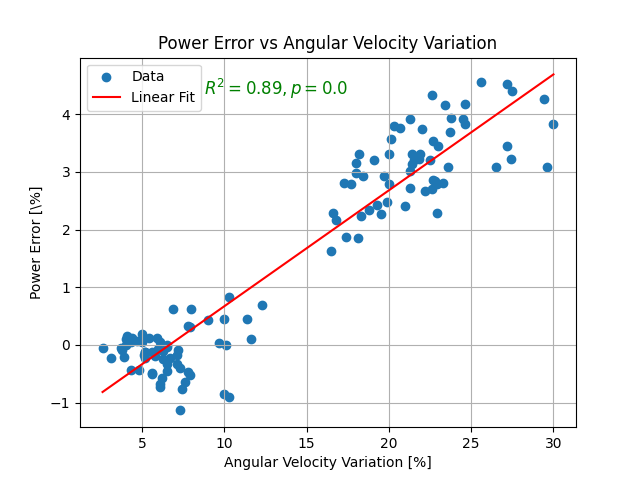} \\
  \caption{Cubic and Linear Fits of Favero Paper dataset}
\end{figure}
Clearly, there is a strong and statistically significant linear correlation between angular velocity variation and error.
Notably, this aggregate dataset combines two categorically distinct datasets,
tests that utilise non-circular (ovalised) chainrings, and tests that utilise circular chainrings. 
The datasets were analysed seperately below. For tests that utilise circular chainrings,
the focus of this paper, there is no statistically significant correlation between
angular velocity variation and error.
\begin{figure}[H]
  \centering
  \includegraphics*[scale=0.5]{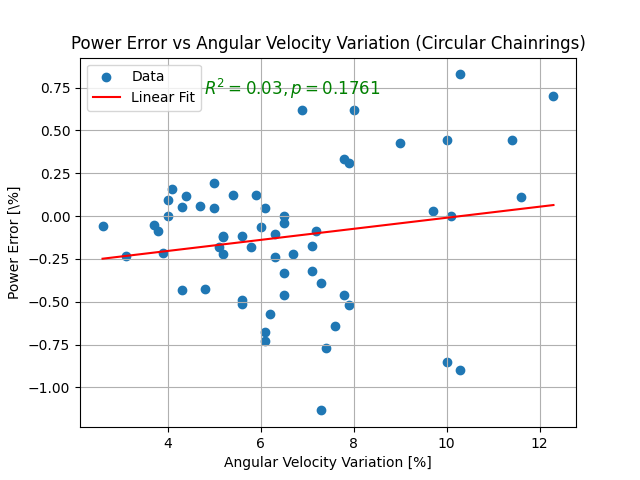}
  \includegraphics*[scale=0.5]{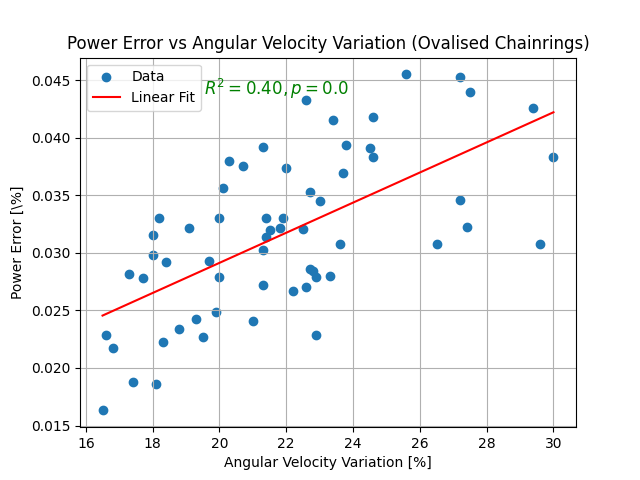}
  \caption{Seperately Considered Linear fit of Circular and Ovalised Chainring Datasets}
\end{figure}
When a Linear fit is applied to \textit{Absolute Error}
for the circular chainring subset of tests, the relationship is small and statistically significant.
There is no individual test
where a statistically significant relationship between angular velcocity variation and absolute error emerges.
This is evident in a randomly selected test ($T2.2.2$), where a statistically significant \textbf{negative error}
exists between the two variables, bucking the trend within the $T2.X.X$ dataset of a positive error relationship.
\begin{figure}[H]
  \centering
  \includegraphics*[scale=0.50]{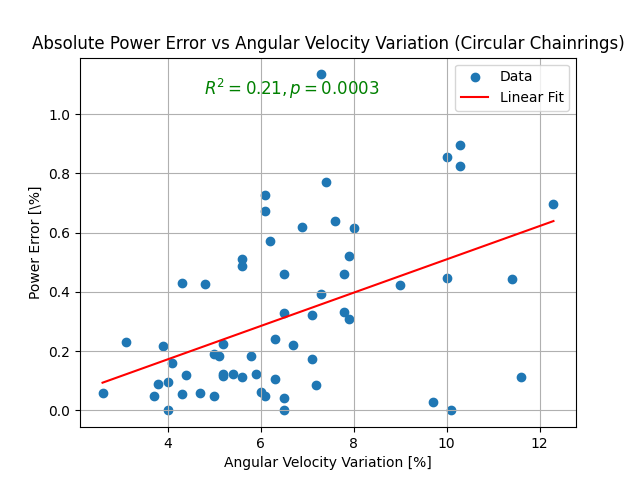}
  \includegraphics*[scale=0.50]{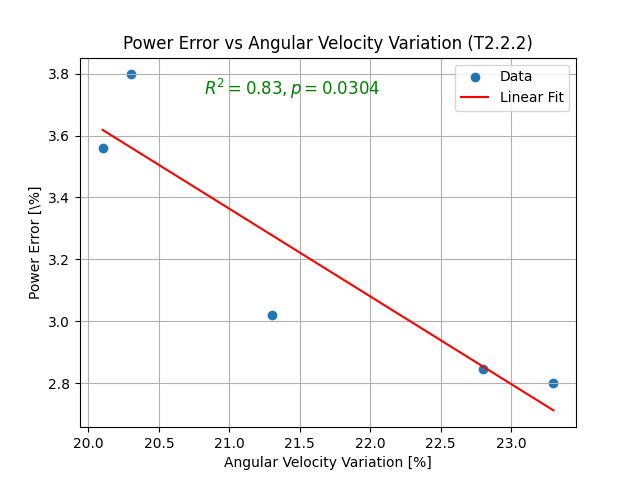}\\
  \caption{Linear Fit of Circular Chainring dataset and $T2.2.2$}
\end{figure}
A causal relationship between Angular Velocity Variation and Power Error 
cannot be claimed. The spurious presense of a strong relationship is the result of Hidden Variables. 
The postulated hidden variable present in this dataset
is phase relationship; ovalised chainrings, by their very nature, introduce a phase relationship
between torque and angular velocity that is unrelated to system dynamics, is not in quadrature with torque,
and thus results in a degree of correlation between AV Variation and Torque.
\subsection{Test Subset}
A subset of the Paper's results,
shown in Table \ref{table:subset} in the Annex, were selected for further analysis.
Specifically selected were those tests that utilise circular chainrings 
and where parameters such as equivelant mass and cadence
were known or could be estimated. This subset of tests
was used to generate a cross-correlation matrix visible
in Table \ref{table:correlations}, with concomitant
p-values visible in Annex A, Table \ref{table:pvalues}. Three synthetic
parameters were generated - 
Absolute Error (E2) $ = \sqrt{E^{2}}$,
Force $ = \frac{IAV}{RPM}$, and FMR $ = \frac{Force}{Mass}$.
\begin{table}[H]
  \centering
  \begin{tabular}{lllllllllll}
    \hline
    & Mass & Grade & RPM & Var & IAV & AAV & Force & FMR & E & E2 \\
    \hline
    Mass & 1.00 & 0.73 & - & - & - & - & - & -0.91 & 0.71 & -0.40 \\
    Grade & 0.73 & 1.00 & -0.49 & - & - & - & 0.39 & -0.65 & 0.63 & - \\
    RPM & - & -0.49 & 1.00 & -0.59 & -0.58 & -0.53 & -0.88 & - & -0.09 & -0.44 \\
    Var & - & - & -0.59 & 1.00 & - & - & 0.49 & - & - & 0.45 \\
    IAV & - & - & -0.58 & - & 1.00 & 0.95 & 0.87 & 0.40 & - & 0.66 \\
    AAV & - & - & -0.53 & - & 0.95 & 1.00 & 0.79 & 0.48 & -0.35 & 0.66 \\
    Force & - & 0.39 & -0.88 & 0.49 & 0.87 & 0.79 & 1.00 & - & - & 0.59 \\
    FMR & -0.91 & -0.65 & - & - & 0.40 & 0.48 & 0.19 & 1.00 & -0.85 & 0.65 \\
    E & 0.71 & 0.63 & - & - & - & -0.35 & - & -0.85 & 1.00 & -0.71 \\
    E2 & -0.40 & - & -0.44 & 0.45 & 0.66 & 0.66 & 0.59 & 0.65 & -0.71 & 1.00 \\
    \hline
    \end{tabular}
    \caption{Subset Correlations}\label{table:correlations}
  \end{table}
Of the non-trivial correlations between dependant and independent variables,
Force to Mass Ratio (FMR) has the strongest correlation with signed error.
Higher FMR results in AAV underestimating power vis à vis IAV. 
There is a moderate and statistically significant relationship between
Absolute Power Error and Mass, Cadence (RPM), Variation, Force and FMR.
\subsection{Conclusion}
The outcome of this section is that the Paper's empirical dataset
is not sufficient to draw any firm, causal conclusions about the relationship between
angular velocity variation and power meter error. For example,
the amount of power itself appears to be a stronger predictor of power error 
than AV Variation. This implies that AV Variation is a side effect of a deeper 
phenomenon that results in power error.\\
The following section will demonstrate that AV Varation, despite being a necessary
condition for power error to exist, is not alone sufficient due to the 
fact that angular velocity is usually in approximate quadrature with net applied torque.
\section{Dynamic Model of a Cyclist}\label{sec:dynamicmodel}
The section will outline the model used for characterising the behaviour of a \textit{Bicycle-Rider System} (BRS)
within a single pedal stroke. 
The behaviour of interest involves small accelerations
and decelerations that produce no net change in velocity over the course of the period. 
Environmental conditions (grade, the relative velocity of the wind, etc) are assumed to remain constant throughout this 
time period - the BRS is in a state of \textit{Dynamic Equilibrium}.
This model will be developed
by first outlining a mechanical and kinematic model of a cyclist under 
conditions of dynamic equilbrium as defined,
and then deriving expressions for acceleration in terms of the parameters of the BRS
and opposing forces.
An Electrical Analogy of the BRS will be introduced, which will
serve as the basis of the modelling which will be performed in Section \ref{sec:modelling}.
\subsection{A Mechanical and Kinematic Model of a Bicycle-Rider System}
The application of force to the pedal of a bicycle
is transferred via a number of mechanical connections
into torque applied at the contact patch of the rear tyre 
to the road, which results in the BRS
applying a net force against external impeding forces.
The BRS has inertia
and drag, and the tyres have rolling resistance. Cycling up a hill
requires a force to be applied against gravity.
A cyclist travelling at $v$ and up a gradient with angle $\theta$ is subject to wind resistance, a quadratic function of the BRS wind velocity difference,
and a force due to gravity:
\begin{equation}
  F_{t} = F_{d} + F_{g} = \frac{1}{2} \rho CdA * (v-v_{wind})^2 + mgsin(\theta)
\end{equation}
Where:
\begin{description}
  \item[$v$] Is the linear velocity of the bicycle-rider system with respect to the ground.
  \item[$v_{wind}$] Is the average velocity of the wind with respect to the ground. 
  \item[$m$] Is the mass of the entire bicycle-rider system.
  \item[$g$] Is the gravitational constant.
  \item[$\rho$] Is air density.
  \item[$Cd$] Is the coefficient of drag of the entire system.
  \item[$A$] Is the frontal surface area of the entire system.
  \end{description}
The average power required to maintain this velocity, assuming velocity remains constant, is given by:
\begin{equation}\label{eq:kinematic}
  P_{avg} = \omega_{avg} \int_{0}^{T} \tau(t) \: dt = v_{avg} \int_{0}^{T} F(t) \: dt = \frac{1}{2} \rho Cd A (v_{avg}-v_{wind})^3 + vmgsin(\theta)
\end{equation}
There are also some frictional forces present resulting from the chain, bearings, and tyres on the road.
Frictional forces tend to be linear with velocity; however there may be some dependance on both $\omega$ and $v$. 
These forces are usually small, and will be ignored for the purposes of this paper.
For the purposes of this paper, a linear approximation will be taken for all resistive 
forces, which gives rise to the following equation of motion:
\begin{equation}
  F = mg \sin(\theta) + k \dot{x} + m \ddot{x}
  \label{eq:motion}
\end{equation}
\subsubsection{Indoor Bicycle Trainers}
Indoor Bicycle Ergometers (also called Bicycle Trainers) have become popular in recent years. 
There are a number of schemes employed to provide resistance to the cyclist. These include
fluid resistance, fan-driven resistance, and magnetic resistance. This paper considers magnetic resistance, the form 
of resistance utilised by the Elite Quobo Trainer in Favero's Paper.
Bicycle Trainers mimic the inertia of a cyclist via the use of a flywheel.
For simple magnetic trainers operating in their linear region, the 
resisitive force generated by Eddy Currents circulating
in the trainer flywheel can be derivied via an application of Lenz's Law:
\begin{equation}
  \epsilon = -N \frac{\partial \phi_{B}}{\partial t}
\end{equation}
Where $\frac{\partial \phi_{B}}{\partial t}$ is generally $\propto \omega_{flywheel}$.
It is possible to linearise the dynamics of this rotating system by equating $\omega_{flywheel}$ with linear
velocity $v$, converting moment of inertia into equivelant mass ($m = kI$),
and considering magnetic resistance to be linearly velocity-dependant.
This produces the same fundamental  of motion as the linear
motion of a cyclist outdoors:
\begin{equation}
  F = C_{m}\dot{x} + M_{eq}\ddot{x}
\end{equation}
\subsection{A Cyclist in Dynamic Equilibrium}
A cyclist's velocity on a bicycle ride, $v(t)$\footnote{Velocity and concomitant quantities are treated as scalar},
will contain some amount of variation.
The angular velocity 
of the pedal, $\omega(t)$, will also vary and,
in a constant gear, $v(t) \propto \omega(t)$\footnote{This is assuming that bicycle gears are perfectly stiff and circular, that force and velocity are seamlessly transferred from the pedal to the rear wheel.}.
This paper distinguishes between two types of variation in velocity:
\begin{enumerate}
  \item \textbf{Oscillations}, which are periodic variations in velocity that are contained within a single pedal stroke.
  \item \textbf{Accelerations}, which are net changes in velocity from the start to the end of a pedal stroke.
\end{enumerate}
If net acceleration over some time period $T$ (the length of a pedal stroke) is zero, then a cyclist is in \textit{Dynamic Equilibrium} over that period.
The mathematical condition for dynamic equilbrium is
\begin{equation}\label{eq:condition}
  \int_{}^{T} \frac{dv(t)}{dt} = \int_{}^{T} \frac{d \omega(t)}{dt} = 0
\end{equation}
Under conditions of dynamic equilibrium, any change in velocity during $T$ (but not over $T$, for there is no change over $T$ by construction)
is purely a product of a previous change in force. 
If a cyclist is in a dynamic equilibrium for period $T$,
it is possible to redefine $v(t)$ as a periodic function (with period $T$)
 by extending the definition of $v(t)$ to $t \in \mathbb{R}$ 
such that $v(t)$ satisfies the condition that, for $a \in \mathbb{Z}, t \in [0,T], v(t) = v(aT+t)$.
$v(t)$ is a smooth function by construction\footnote{$v(t)$ is smooth by construction due to the fact that, for $a \in \mathbb{Z}$, the limit of $v(aT + t)$ as $t \to T$ (and $\therefore aT  + 1 \to (a+1)T) = v((a+1)T$ by virtue of the extended definition of $v(t)$}, and finite due to the fact that
it represents a real quantity of only finite values. 
Given $v(t)$ is smooth and finite, bounded variation can be asserted
which is the condition of the Dirichlet–Jordan test, and the Fourier Series $S_{n}(v(t))$ converges to $v(t)$.
  \subsubsection{Force and Acceleration}
  Force and Acceleration are not necessarily smooth time-varying quantities, however in physical systems they are finite-valued. 
  A differential equation for acceleration can be derived by expressing force as a Fourier Series\footnote{As will be discussed in the following section, it is empirically true that force can be expressed as a Fourier Series and can be well-approximated with a small number of terms}
  and utilising a linear approximation of opposing forces.
  \begin{equation}
    \frac{dv(t)}{dt} = \frac{1}{m}(F_{0} + \sum_{n=1}^{\infty}F_{n}\sin(2\pi nt + \phi_{n}) - kv - mg\sin(\theta))
  \end{equation}
  The periodicity of force application coincides with the periodicity of velocity,
  since changes in velocity are entirely driven by the periodic application of force (and the monotonicly
  increasing impeding function)\footnote{This claim is not proven explicitly, however it is self evidently true in the case of the electrical analogue introduced later}.
  Because the gear-ratio is assumed to remain constant, for $t \in [0,T], v \propto \omega$ and $\tau \propto F$\footnote{The relationship between
  $\tau \& F$ and $\omega \& v$ is through power balance $\tau \times \omega = F * v$; $\omega(t) = k_{0} v(t)$ and $\tau(t) = k_{1} F(t)$}.
  \subsection{Power, Averaging, and Oscillations in Velocity and Force}
  A primary aim of this paper is to explore how oscillations in 
  linear or angular velocity might contribute to differences in 
  the power produced and measured in a state of dynamic equilibrium
  when differing techniques for computing power, Average Angular Velocity (AAV) and Instantenous Angular Velocity (IAV),
  are employed. There are two theoretical sources of error that might result from averaging angular velocity.
  \subsubsection{Cross-Correlation Error}
  Torque and Angular Velocity can be described as a quantities
  that continuously vary in time - $\tau(t)$ and $\omega(t)$ - from which it is possible to calculate $P(t) = \tau(t) \times \omega(t)$. 
  If we are interested in the average power, labelled $\overline{P}$, over some period $T$ we will not
  arrive at a precise result by considering the product of the averages of the individual components, 
  $\overline{\tau}*\overline{\omega}$, due to the precense of \textbf{cross-correlation}. 
  For two differentiable functions, $f$ and $g$, it is generally the case that:
  \begin{equation}
    \int f(x)g(x) \: dx \neq \int f(x) \: dx \int g(x) \: dx
    \label{eq:crosscorrelation}
  \end{equation}
  Cross-Correlation Error could be either positive or negative, depending on the sign
  of the cross-correlation between the two functions under consideration. 
  \subsubsection{Non-Conservative Forces and Jensen's Inequality}
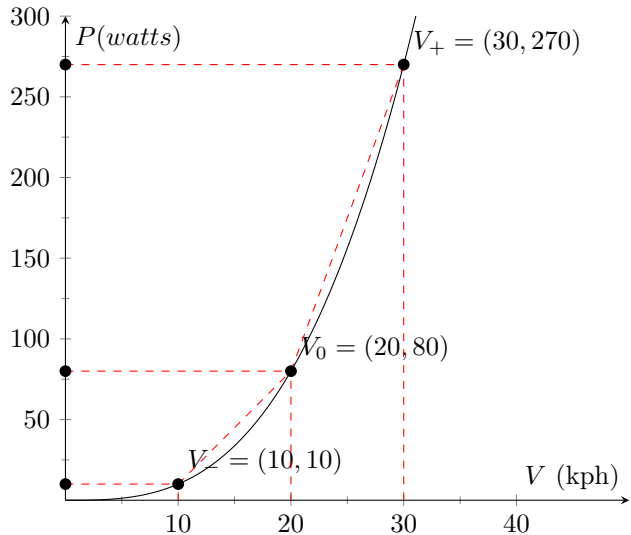
\begin{figure}[H]
    \centering
  \begin{tikzpicture}
    \begin{axis}[
      xlabel={$V$ (kph)},
      ylabel={$P (watts)$},
      axis lines=middle,
      xmax=50, xmin=0,
      ymax=300, ymin=0,
      xtick={0,10,20,30,40},
      xticklabels={$0$,$10$,$20$,$30$,$40$},
      minor tick num=1,
      major grid style={line width=.2pt,draw=gray!50},
      minor grid style={line width=.2pt,draw=gray!20},
      width=9cm,
      height=8cm,
      ]
      \addplot[black, samples=100, domain=0:40] {0.01*x^3} node[right, pos=0.65] {$P=0.01V^3$};
      \addplot[mark=*] coordinates {(10, 10)} node[above right] {$V_{-}=(10,10)$};
      \addplot[mark=*] coordinates {(20, 80)} node[above right] {$V_{0}=(20,80)$};
      \addplot[mark=*] coordinates {(30, 270)} node[above right] {$V_{+}=(30,270)$};
          \addplot[dashed, red] coordinates {(10, 10) (20, 80)};
          \addplot[dashed, red] coordinates {(20, 80) (30, 270)};
          \addplot[dashed, red] coordinates {(0, 10) (10, 10)};
          \addplot[dashed, red] coordinates {(0, 80) (20, 80)};
          \addplot[dashed, red] coordinates {(0, 270) (30, 270)};
          \addplot[dashed, red] coordinates {(10, 0) (10, 10)};
          \addplot[dashed, red] coordinates {(20, 0) (20, 80)};
          \addplot[dashed, red] coordinates {(30, 0) (30, 270)};
          \addplot[mark=*] coordinates {(0, 270)} node[above right] {};
          \addplot[mark=*] coordinates {(0, 80)} node[above right] {};
          \addplot[mark=*] coordinates {(0, 10)} node[above right] {};
    \end{axis}
  \end{tikzpicture}
  \caption{An illustration of how Jensen's Inequality may contribute to reported power error.}\label{fig:jensen}
  \end{figure}

  There is a third order polynomial relationship that characterises the relationship between velocity and power.
  A realistic example of this relationship is shown in Figure \ref{fig:jensen}.
  This relationship between velocity and the power required to maintain
  that velocity[\ref{eq:kinematic}] means that oscillations or variations about an average velocity value
  result in higher average power requirements for a given average speed, and the greater the variance 
  in velocity, the greater the average power requirements for the same speed. 
  This is due to Jensen's inequality; in Figure \ref{fig:jensen}, the size of the projection of the secant line
  connecting $V_{0}$ and $V_{+}$ onto the Power-Axis is greater that the same projection of the secant
  line connecting $V_{0}$ and $V_{-}$.
  For the purposes of illustration, consider the power computed
  by averages and properly for a cyclist travelling 20km during 
  the course of one hour in an oscillatory fashion. 
  Specifically, the cyclist travels at At 30kph ($270W$) for 30 minutes
  and then 10kph($10W$) for the remainder. AAV would predict an average
  power of $80W$ (20kph), whereas IAV would produce an average power of $140W$ (neglecting accelerations).
  \subsection{A Mathematical Model of the Pedal Stroke}\label{ss:pedalstroke}
  The cycling pedal stroke is characterised by the application of torque by
  the cyclist. The application of torque is quasiperiodic; that is,
  there is a regular variation within pedal strokes, but also some level
  of variation over longer time periods (due to changes in gear, gradient,
  decisions to accelerate or decelerate). This section shall characterise the variation
  in force application within a single pedal stroke.
\subsubsection{An Empirical Model of the Pedal Stroke}
  Bini \& Carpes, 2014\cite{bini2014biomechanics} measured the force exerted by a certain cyclist on a pedal throughout the course of a single pedal stroke,
  with results presented in Figure \ref{fig:pedalstroke}. This pedal stroke will be characterised 
  and used as the basis for simulation.
  The pedal stroke developed is one particular pedal stroke, 
and there likely some amount of variation between cyclists, 
as well as variation across different conditions and with fatigue.
In spite of this fact, this pedal stroke does contain non-idealities
which are potential contributing factors to power error, such as an imbalance
between left and right torque application\footnote{The results of this
paper could be more strongly claimed with a thorough empirical model of the pedal stroke}.\\
Temporal variation in torque was analysed in terms of its frequency content\footnote{Using a Discrete Fourier Transform (DFT)},
and harmonics are presented in terms of their amplitude and phase angle in Table \ref{table:harmonics}.
Harmonics above the fifth were omitted for the sake of brevity.\\
Total Harmonic Distortion (THD) was calculated with respect to the DC content of the pedal stroke.
\begin{equation}
  THD (\%) = \frac{\sqrt{\sum_{i=i}^{5}V_{i}}}{V_{0}}
\end{equation}
\begin{figure}[H]
  \centering
  \begin{tikzpicture}
    \begin{axis}[
      axis lines=left, 
      axis line style={-}, 
      xmin=-10, xmax=370, 
      ymin=-0.5, ymax=2.5, 
      width=12cm, height=10cm, 
      axis on top, 
      clip=false, 
      xlabel={Angle (degrees)}, 
      xtick={0,30,...,360},
    ]
        
    \pgfmathsetmacro\Aone{75}
    \pgfmathsetmacro\sigmaone{25}
    \pgfmathsetmacro\inertia{5}
    \pgfmathsetmacro\basicv{20}
    \pgfmathsetmacro\offset{10}

    \pgfmathsetmacro\Atwo{5}
    \pgfmathsetmacro\sigmatwo{25}
  
    \addplot[domain=0:360, samples=1000, black, smooth] {1.0+2*(0.05*cos(x-121.386)+0.42*cos(2*x+157.439)+0.039*cos(3*x+129.510)+0.064*cos(4*x+45.443)+0.023*cos(5*x-118.745))};
    \addlegendentry{Net Torque};
    \addplot[domain=0:360, samples=1000, red, smooth] {0.521+2*(0.545*cos(x-99.993)+0.205*cos(2*x+158.783)+0.069*cos(3*x+89.984)+0.021*cos(4*x+29.845)+0.007*sin(5*x-43.564))};
    \addlegendentry{Left Torque};
    \addplot[domain=0:360, samples=1000, blue, smooth] {0.479+2*(0.499*cos(x+82.08)+0.220*cos(2*x+156.185)+0.046*cos(3*x-123.439)+0.044*cos(4*x+52.782)+0.022*cos(5*x-137.366))};
    \addlegendentry{Right Torque};

  \end{axis}
  \end{tikzpicture}
  \caption{Variations in Applied Torque}\label{fig:pedalstroke}
\end{figure}
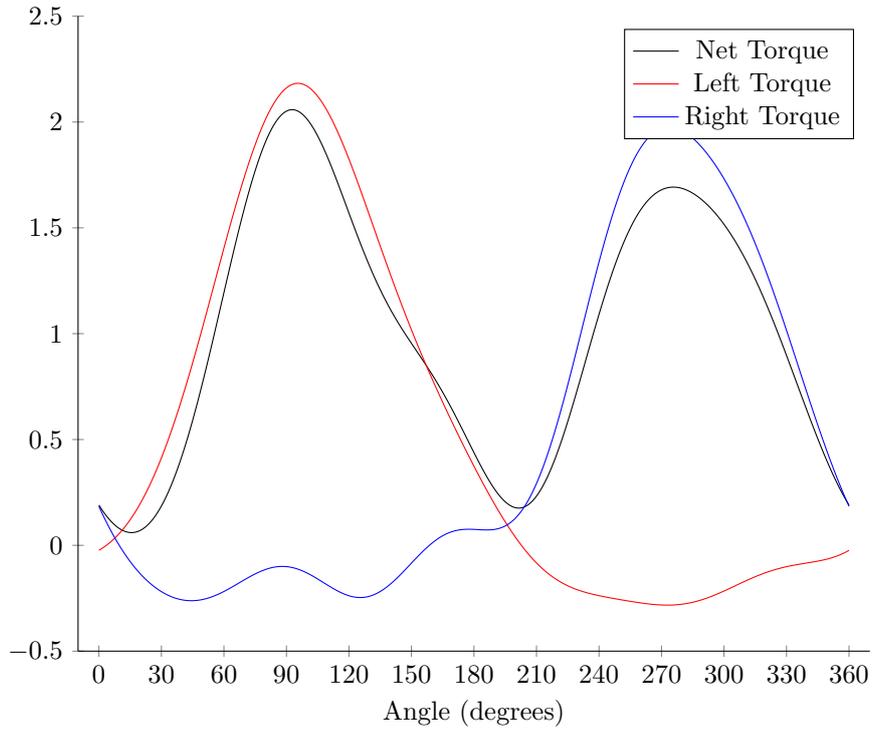
\begin{table}[H]
  \centering
  \begin{tabular}{|c|c|c|c|}
  \hline
  & Net Torque & Left Torque & Right Torque \\
  \hline
  DC Component & $1.000 \angle -0.000$ & $0.521 \angle -0.000$ & $0.479 \angle -0.000$ \\
  1st Harmonic & $0.099 \angle -121.386$ & $1.089 \angle -99.993$ & $0.997 \angle 82.088$\\
  2nd Harmonic & $0.849 \angle 157.439$ & $0.410 \angle 158.783$ & $0.439 \angle 156.185$\\
  3rd Harmonic & $0.079 \angle 129.510$ & $0.137 \angle 89.984$ & $0.091 \angle -123.439$\\
  4th Harmonic & $0.127 \angle 45.443$ & $0.042 \angle 29.845$ & $0.088 \angle 52.782$\\
  5th Harmonic & $0.045 \angle -118.745$ & $0.014 \angle -43.564$ & $0.044 \angle -137.366$\\
  \hline
  $THD (\%)$ & 193\% & 284\% & 287\%\\
  \hline
  \end{tabular}
  \caption{Normalised Harmonic Content of a Typical Pedal Stroke}\label{table:harmonics}
\end{table}
It should be noted that there is an upper limit on the amount of THD that could realistically be present in a pedal stroke,
due to the fact that the freehub ratchet prevents the application of negative torque to the rear wheel. Future
work could involve determining an upper limit on power error based on this physical limitation, as opposed a single empirical
model of a pedal stroke.
\subsubsection{Induced Variations in Angular Velocity}
Variations in net torque give rise to variations in Angular Velocity.
This is represented below considering
only the dominant harmonic of angular velocity; observe the phase relationship between torque and angular velocity. That is,
angular velocity \textit{lags} the application of torque
due to the fact that torque variation \textit{induces} 
angular velocity variation. For reasons that will soon become apparent, this phase angle is strictly
bounded to $0 < \theta < \frac{\pi}{2}$ under conditions of dynamic equilbrium.
  \begin{figure}[H]
    \centering
    \begin{tikzpicture}
      \begin{axis}[
        axis lines=left, 
        axis line style={-}, 
        xmin=-10, xmax=370, 
        ymin=-0.5, ymax=2.5, 
        width=12cm, height=10cm, 
        axis on top, 
        clip=false, 
        xlabel={Angle (degrees)}, 
        xtick={0,30,...,360},
      ]
          
      \pgfmathsetmacro\Aone{75}
      \pgfmathsetmacro\sigmaone{25}
      \pgfmathsetmacro\inertia{5}
      \pgfmathsetmacro\basicv{20}
      \pgfmathsetmacro\offset{10}
  
      \pgfmathsetmacro\Atwo{5}
      \pgfmathsetmacro\sigmatwo{25}
    
      \addplot[domain=0:360, samples=1000, black, smooth] {1.0+2*(0.05*cos(x-121.386)+0.42*cos(2*x+157.439)+0.039*cos(3*x+129.510)+0.064*cos(4*x+45.443)+0.023*cos(5*x-118.745))};
      \addlegendentry{Net Torque ($\tau(\theta)$)};
      \pgfplotsextra{
          \pgfmathsetmacro{\avgvalue}{1.0}
          \draw[red, dashed] (axis cs: 0,\avgvalue) -- (axis cs: 360,\avgvalue) node[right, pos=0.95, red] {\tiny{Average Force}};
      }
      
    \pgfmathsetmacro\average{35}
    \addplot[domain=0:360, samples=1000, red, smooth] {1.0 + 0.05*cos(2*x+157.439 - 90)};
      \addlegendentry{Angular Velocity ($\omega(\theta)$)};
      \node[rotate=90] at (-50,105) {Applied Torque / Angular Velocity};
      \end{axis}
    \end{tikzpicture}
    \caption{Variations in Applied Torque and Angular Velocity}
  \end{figure}
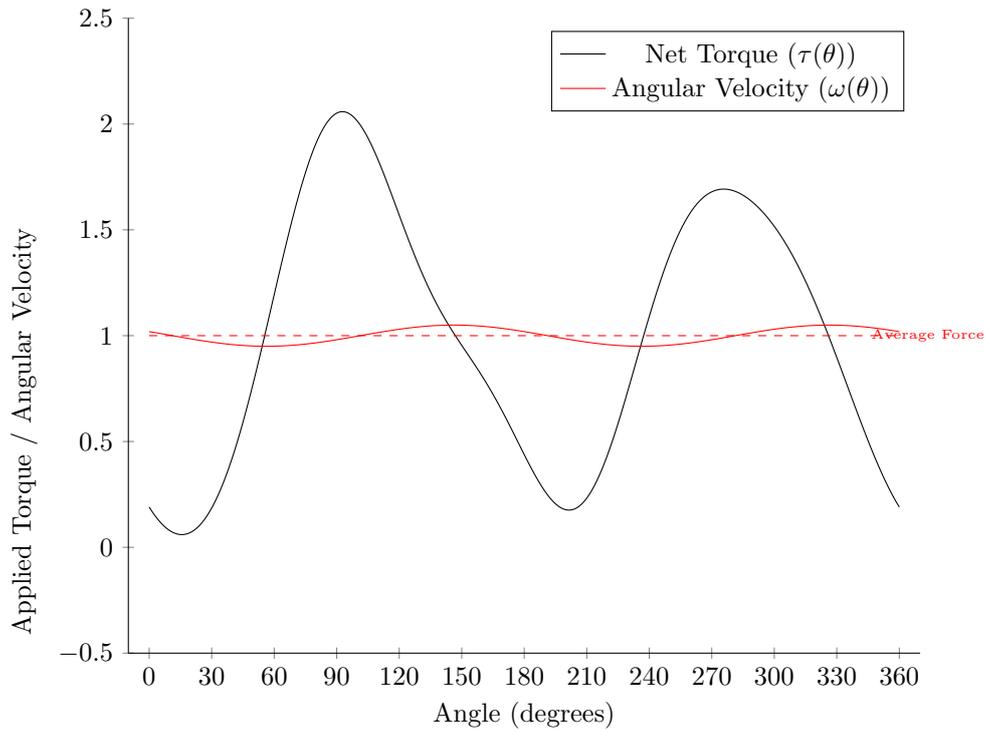
  The following section will develop the mathematical relationship between applied torque and induced angular velocity.
  relates to applied torque.
  \subsection{System Modelling}
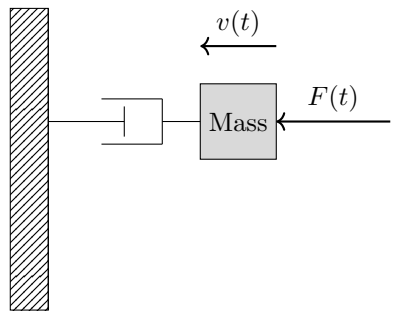
\begin{figure}[H]
    \centering
    \begin{tikzpicture}
    \draw (0,2) -- (0,-2);
    \draw[pattern=north east lines, pattern color=black] (-0.5,2) rectangle (0,-2);
    \draw[fill=gray!30] (2,0) rectangle node[text width=2cm,align=center] {Mass} (3,1);
      
      \draw[->,thick] (4.5,0.5) -- node[above]{$F(t)$} (3,0.5);
      \draw[->,thick] (3,1.5) -- node[above]{$v(t)$} (2,1.5);
    \draw (0,0.5) -- (1,0.5);
    \draw (1,0.3) -- (1,0.7);
    \draw (0.7,0.2) -- (1.5,0.2);
    \draw (0.7,0.8) -- (1.5,0.8);
    \draw (1.5,0.2) -- (1.5,0.8);
    \draw (1.5,0.5) -- (2.0,0.5);

    \end{tikzpicture}
    \caption{Mass-Damper Model of a Cyclist}\label{fig:massdamper}
  \end{figure}

\begin{figure}[H]
    \centering
    \begin{circuitikz}
    \draw (0,0) to[V, v=$V$] (0,2)
                to[V, v<=$Vg$, invert] (0,4)
                to[L, l=$L$] (4,4)
                to[R, l=$R$] (4,0)
                -- (0,0);
                \draw (-0.3,2.2) node[label={[font=\footnotesize, below]:\textbf{$V_{x}$}}]{};
                \draw (0,1.0) -- ++(0,0.5) coordinate (currentarrowstart);
                \draw[-{Latex[width=2mm]}] (currentarrowstart) -- ++(0,0.5) node[above, right] {\textbf{$I_{x}$}};
\end{circuitikz}
\caption{General Circuit Model of a BRS}\label{fig:generalcircuit}
\label{fig:cyclistcircuit}
\end{figure}
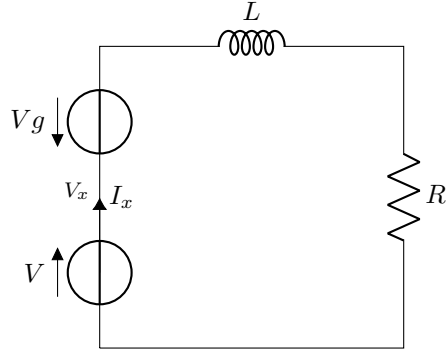

  A cyclist under dynamic equilibrium faces a net impeding force that is a function
  of their velocity. If a linear approximation of these resistive
  forces is used, then the BRS to be modelled as a mass-damper system as shown in Figure \ref{fig:massdamper}.
Sources of impedance (to movement)
can be replaced with their electrical analogues (Inductance
for Mass, and Resistance for Friction and Wind Resistance). 
Extending the analogy, the Force applied by the system
(which is transferred from pedal force) is represented
by Voltage, and the resultant Current represents Velocity.
This is represented in Figure \ref{fig:generalcircuit} below - 
$V$ represents the applied force, and $V_{g}$ represents
the constant opposing force due to gravity.
The value of resistance is as follows
\begin{equation}\label{eq:resistance}
  R = \frac{F_{d}(v)}{v}
\end{equation}
Consider $v(t)$, representing the force applied by the cyclist over time, to be a Fourier Series:
\begin{equation}
  v(t) = v_{0} + \sum_{n=1}^{\infty} v_{n} \sin(2\pi nt + \phi_{n})
\end{equation}
In the frequency domain, the Reactive Impedance seen by the $n_{th}$ harmonic is given by:
\begin{equation}
  X_{n} = j \omega nL
\end{equation}
Taking a phasor representation of that particular harmonic (phasor quantities are RMS, and denoted with capital letters),
the $n_{th}$ harmonic current induced in the circuit resulting from the applied harmonic voltage is given as:
\begin{equation}
  I_{x,n} = \frac{V_{x,n}}{X_{n} + R}
\end{equation}
By the principle of superposition, the total current in the circuit 
is the sum the current contributions of individual harmonics, so total current in the circuit can also be expressed as a Fourier Series:
\begin{equation}
  i(t) = i_{0} + \sum_{m=1}^{\infty} i_{m} \sin(2\pi mt + \phi_{m})
\end{equation}
Now consider the computation of the power dissapated by the resistor over 
one period of the fundamental frequency ($T$). First note the following, via an application
of the general fact established in Equation \ref{eq:crosscorrelation}:
  \begin{align*}
    P_{avg} =  \frac{1}{T} \int_{0}^{T} V(t) I(t) \: dt \neq \frac{1}{T} \int_{0}^{T} V(t) \: dt * \frac{1}{T} \int_{0}^{T} I(r) \: dt
  \end{align*}
  If there is in fact greater or lesser power produced in reality (as measured by IAV)
than is computed via AAV, this negative \textit{Power Error} represents power
disappated into the wind and to other frictional forces by the cyclist due to oscillations\footnote{Positive Power Error would indicate external forces performing harmonic work on the BRS}.
  The following section will characterise harmonic losses in RL circuits in terms of the interaction between the various harmonics.
  \subsubsection{Harmonic Losses in RL Circuits with multiple harmonics and a DC Offset}\label{sec:HarmonicLosses}
Over a single period (normalised to 1 second), the energy dissapated by a resistor in an RL circuit with periodic 
and incoherant voltage source is given by:
  \begin{equation}
    \label{eq:harmonics}
    \int_{0}^{1} v(t)*i(t) \: dt = \int_{0}^{1} (v_{0} + \sum_{n=1}^{\infty} v_{n} \sin(2 \pi nt + \phi_{n}) ) * (i_{0} + \sum_{m=1}^{\infty} i_{m} \sin(2\pi mt + \psi_{m}) ) \: dt = v_{0}i_{0} \pm \epsilon
  \end{equation}
  Where $v_{0}i_{0}$ is the power dissapated by the DC components of voltage
  and current, and $\epsilon$ is the power dissapated by harmonics. Note the following lemmas for $n,m \in \mathbb{N}$ where $n \neq m$ and $\phi_{n}, \phi_{m} \in \mathbb{R}$:
  \begin{equation} \label{lemma:sincancellation}
    \int_{0}^{1} \sin(2\pi nt + \phi) \: dt = 0
  \end{equation}
  \begin{equation}
    \label{lemma:sameharmonics}
    \int_{0}^{1} \sin(2\pi nt + \phi) * \sin(2\pi nt + \phi + \frac{\pi}{2}) \: dt = 0
  \end{equation}
  \begin{equation}
    \label{lemma:differentharmonics}
    \int_{0}^{1} \sin(2\pi nt + \phi_{n}) * \sin(2\pi mt + \phi_{m}) \: dt = 0
  \end{equation}
  The integrad of Equation \ref{eq:harmonics} can be expanded binomially, yielding five distinct terms:
  \begin{equation} \label{eq:dc}
    \int_{0}^{1} v_{0} * i_{0} \: dt = 0
  \end{equation}
  \begin{equation} \label{eq:cancellation1}
    \int_{0}^{1} ( \sum_{n=1}^{\infty} v_{n} \sin(2 \pi nt + \phi_{n}) ) * (i_{0}) \: dt = 0
  \end{equation}
  \begin{equation} \label{eq:cancellation2}
    \int_{0}^{1} (v_{0})*(\sum_{k=1}^{\infty} i_{k} \sin(2 \pi kt + \psi_{k}) ) \: dt = 0
  \end{equation}
  \begin{equation} \label{eq:sameharmonics}
    \int_{0}^{1} \sum_{n=1}^{\infty} (v_{n} \sin(2 \pi nt + \phi_{n}))*(i_{n} \sin(2 \pi nt + \psi_{n}) ) \: dt
  \end{equation}
  \begin{equation}\label{eq:diffharmonics}
    \int_{0}^{1} \sum_{n=1}^{\infty} \sum_{k=1, k \neq n}^{\infty} (v_{n} \sin(2 \pi nt + \phi_{n}))*( i_{k} \cos(2 \pi kt + \psi_{k}) ) \: dt = 0
  \end{equation}
  By the fact of of Lemma \ref{lemma:sincancellation}, Equations \ref{eq:cancellation1}
  and \ref{eq:cancellation2} evaluate to zero. By the fact of Lemma \ref{lemma:differentharmonics},
  Equation \ref{eq:diffharmonics} evaluates to zero. By the fact of Lemma \ref{lemma:sameharmonics},
  Equation \ref{eq:sameharmonics} evaluates to zero when harmonics of the same order are in quadrature.
  Therefore, the only contribution to $\epsilon$ derives from harmonics of the same order that are not in quadrature.
  This can be computed as follows:
  \begin{equation}\label{eq:powererror}
    \epsilon = \sum_{n=1}^{\infty} \mathbb{R}(V_{x,n}*I_{x,n}) = \sum_{n=1}^{\infty} \frac{|V_{x,n}|^2}{|j\omega nL + R|} \cos(\tan^{-1}(\frac{R}{j\omega L}))
  \end{equation}
  Note that $\epsilon$ is strictly positive, and this is due to the fact that resistance
  must take on a positive value.
  \subsubsection{Unilateral Torque Measurement} \label{sec:unilateral}
\begin{figure}[H]
      \centering
      \begin{circuitikz}
      \draw (0,0) to[V, v=$V_{l}$] (0,2)
                  to[V, v=$V_{r}$] (0,4)
                  to[L, l=$L$] (4,4)
                  to[R, l=$R$] (4,0)
                  -- (0,0);
                  \draw (-0.3,2) node[label={[font=\footnotesize, below]:\textbf{$V_{l}$}}]{};
                  \draw (-0.3,4.0) node[label={[font=\footnotesize, below]:\textbf{$V_{x}$}}]{};
                  \draw (0,4) -- ++(0.5,0) coordinate (currentarrowstart);
                  \draw[-{Latex[width=2mm]}] (currentarrowstart) -- ++(0.5,0) node[above, right] {\textbf{$I_{x}$}};
  \end{circuitikz}
  \caption{Circuit Model of a Cyclist Producing Unilateral Torque}\label{fig:unilateral}
  \end{figure}
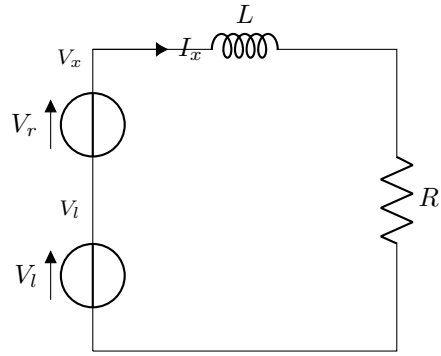

  Torque application is unilateral\footnote{That is, torque is applied by both feet to the crankarms independently, which produces a resultant net torque on the crankarm}, 
  and may be measured unilaterally,
  however angular acceleration is a function 
  of net torque. Extending the electrical analogy,
  power produced unilaterally can be expressed as two individual voltage sources, each with a characteristic harmonic profile:
  \begin{align*}
    v_{l}(t) = v_{l,0} + \sum_{n=1}^{\infty} v_{l,n}\sin(2\pi nt + \phi_{n})\\
    v_{r}(t) = v_{r,0} + \sum_{n=1}^{\infty} v_{r,n}\sin(2\pi nt + \psi_{n})
  \end{align*}
  As shown in Figure \ref{fig:unilateral}, voltage sources are arranged head-to-tail, and result in a certain 
  node voltage, which represents the net torque/voltage; it is the net voltage at this note which induces a current in the circuit.
  The net voltage can then be expressed as a time-domain summation:
  \begin{equation}
    v_{x}(t) = v_{l}(t) + v_{r}(t) = v_{x,0} + \sum_{n=1}^{\infty} v_{x,n}\sin(2\pi nt + \delta_{n})
  \end{equation}
  Phasor Representation for a particular harmonic ($n$) is possible for for the Left($V_{l,n}$), Right($V_{r,n}$) and Net($V_{x,n}$) Voltages, and for net current $I_{x,n}$.
  Unilateral power error can be computed as follows:
  \begin{equation}\label{eq:unilateralerror-l}
    P_{l} = \sum_{n=1}^{\infty} \mathbb{Re}(V_{l,n} * I_{x,n})\\
  \end{equation}
  \begin{equation}\label{eq:unilateralerror-r}
    P_{r} = \sum_{n=1}^{\infty} \mathbb{Re}(V_{r,n} * I_{x,n})
  \end{equation}
  In Table \ref{table:harmonics}, it is clear that the magnitude of the first harmonic of torque as measured on the left and right
  crankarms is substantially higher than the magnitude of net torque; it is also
  obvious that there is phase displacement of left and right 
  torque with respect to net torque. We might thus predict the precense of spurious
  power error if unilateral power was measured.
  \subsubsection{A Tangential Comment on Dynamic Inequilibrium}
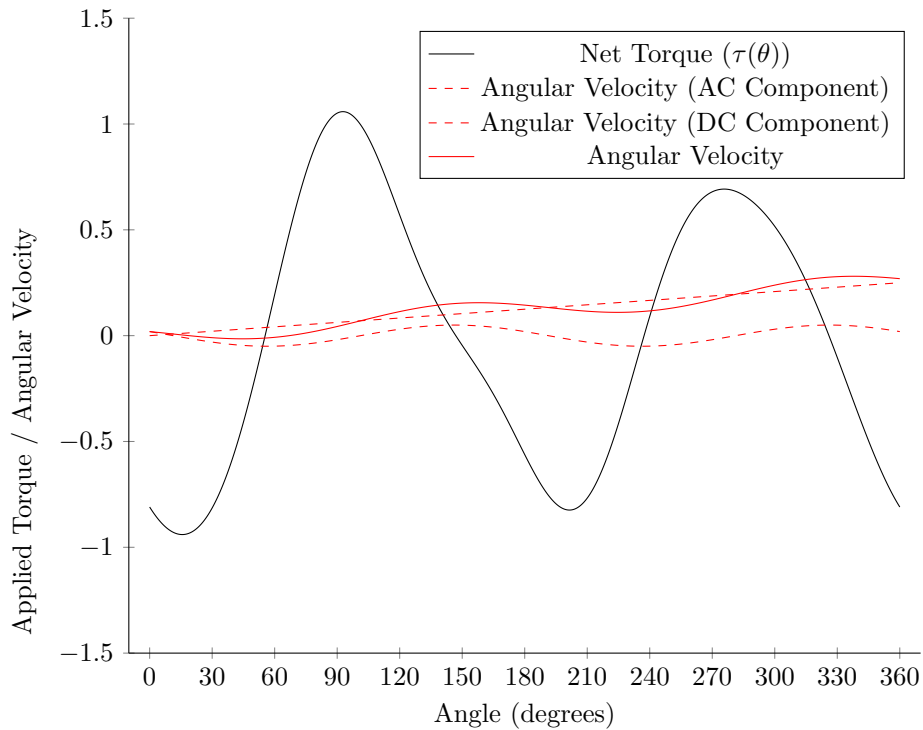
\begin{figure}[H]
    \centering
    \begin{tikzpicture}
      \begin{axis}[
        axis lines=left, 
        axis line style={-}, 
        xmin=-10, xmax=370, 
        ymin=-1.5, ymax=1.5, 
        width=12cm, height=10cm, 
        axis on top, 
        clip=false, 
        xlabel={Angle (degrees)}, 
        xtick={0,30,...,360},
      ]
          
      \pgfmathsetmacro\Aone{75}
      \pgfmathsetmacro\sigmaone{25}
      \pgfmathsetmacro\inertia{5}
      \pgfmathsetmacro\basicv{20}
      \pgfmathsetmacro\offset{10}
  
      \pgfmathsetmacro\Atwo{5}
      \pgfmathsetmacro\sigmatwo{25}
    
      \addplot[domain=0:360, samples=1000, black, smooth] {2*(0.05*cos(x-121.386)+0.42*cos(2*x+157.439)+0.039*cos(3*x+129.510)+0.064*cos(4*x+45.443)+0.023*cos(5*x-118.745))};
      \addlegendentry{Net Torque ($\tau(\theta)$)};

    \pgfmathsetmacro\average{35}
    \addplot[domain=0:360, samples=1000, red, dashed] {0.05*cos(2*x+157.439 - 90)};
    \addplot[domain=0:360, samples=1000, red, dashed] {(1/360 * 0.25 * x)};
    \addplot[domain=0:360, samples=1000, red, smooth] {0.05*cos(2*x+157.439 - 90) + (1/360 * 0.25 * x)};
      \addlegendentry{Angular Velocity (AC Component)};
      \addlegendentry{Angular Velocity (DC Component)};
      \addlegendentry{Angular Velocity};
      \node[rotate=90] at (-50,105) {Applied Torque / Angular Velocity};
      \end{axis}
    \end{tikzpicture}
    \caption{Variations in Applied Torque and Angular Velocity - Dynamic Inequilibrium}\label{fig:dynamicinequilib}
  \end{figure}

  Under conditions of Dynamic Inequilibrium, initial 
  velocity does not equal terminal velocity.
  This introduces a DC Term $F(t)$ which could take an arbitrary form,
  represented in Figure \ref{fig:dynamicinequilib}.
  The resultant error essentially derives from the extent to which the DC component deviates from linearity.
\section{Model Validation and Extrapolation}\label{sec:modelling}
The purpose of this section is to numerically test the 
model developed in Section \ref{sec:dynamicmodel}
against the subset of results from Favero's Paper as shown in 
Table \ref{table:subset}, and then to use the model
to predict scenarios where Power Error might occur
beyond the scenarios tested experimentally.
The model essentially encodes the dynamic model of a cyclist developed above, and 
computes power error by taking the following inputs:
\begin{description}
  \item[$v$] as the linear velocity of the bicycle-rider system with respect to the ground.
  \item[$v_{wind}$] as the average velocity of the wind with respect to the ground. 
  \item[$m$] as the mass of the entire bicycle-rider system. This value was estimated at 75kg for a cyclist riding outdoors, and 3kg for an indoor trainer.
  \item[$g$] as the gravitational constant.
  \item[$\frac{1}{2}\rho CdA$] is used to determine drag. For a cyclist riding outdoors, this was estimated to be $0.25m^{2}$, and a higher value of $0.5m^{2}$ was used for the indoor trainer.
  \item[$Cadence$] is the rate at which the pedal performs a full revolution about the bottom bracket.
  \item[$Harmonic Profile$] as per Table \ref{table:harmonics}
  \end{description}
The model computes Resistance according to Equation \ref{eq:resistance}.
AAV is computed according to Equation \ref{eq:kinematic}.
Power Error is computed according to Equation \ref{eq:powererror},
and Unilateral Power Errors for left and right are computed according to Equations
\ref{eq:unilateralerror-l} and \ref{eq:unilateralerror-r}. Variation is computed according to:
\begin{equation}
  Var (\%) = \frac{\sqrt{\sum_{n=1}^{\infty} (2 * V_{x})^2}}{v}
\end{equation}
The developed model is sensitive to input parameters, many of which are unknown
(such as actual mass, actual harmonic profile, actual drag). 
This may explain the deviation between model results and experimental data.
\subsubsection{Model Fidelity}
For each scenario enumerated in Table \ref{table:scenariodefs},
a velocity was chosen for $v$ of Table \ref{table:scenarioresults} (to one DP of precision) that most
closely matched the resultant power value from $P_{target}$ of Table \ref{table:scenariodefs}.
It should be noted that simulated AV Variation was generally lower 
than measured. This could be due to a number of experimental factors such as measurement noise,
or deviation from Dynamic Equilibrium. There were no conditions 
where Power Error ($\epsilon_{x}$) was significant ($>0.25\%$),
however unilateral power error reached as high as $0.42\%$. 
It can be concluded that, in high inertia scenarios, any power 
error indicated in Favero's Paper cannot be attributed to the interaction
of harmonics of torque and angular velocity within the pedal stroke.
\begin{table}[H]
  \centering
  \begin{tabular}{|c|c|c|c|c|c|c|c|}
    \hline
    \# & Name      & Mass (kg) & $\frac{1}{2} \rho CdA$ $(m^{2})$ & Cadence (rpm) &  $P_{target}$ (W) & $Var_{target}$ (\%) & $\epsilon_{target}$ \\
    \hline
    & Outdoors & \multicolumn{5}{|c|}{}\\
    \hline
    1 & Flat Ground, 100rpm & 75 & 0.25 & 100& 192.08 & 6.2 & 0.04\\
    2 & Flat Ground, 90rpm  & 75 & 0.25 & 90 & 212.90 & 6.7 & 0.10\\
    3 & 5\% Grade\tablefootnote{No cadence value was provided in Favero's Paper, so an average cadence of 60rpm was assumed}           & 75 & 0.25 & 60 & 295.16 & 8.1 & 0.48\\
    \hline
    & Trainer\tablefootnote{Trainer resistance is assumed to be higher than the equivelant outdoor resistance} & \multicolumn{5}{|c|}{}\\
    \hline
    4 & 70\% FTP 90rpm      & 4 & 0.5 & 90 & 230.26 & 4.8 & -0.36\\
    5 & 70\% FTP 110rpm     & 4 & 0.5 & 110& 213.18 & 4.2 & -0.22\\
    6 & 95\% FTP 90rpm      & 4 & 0.5 & 90 & 287.24 & 6.3 & -0.51\\
    7 & 95\% FTP 70rpm      & 4 & 0.5 & 70 & 270.14 & 8.0 & -0.86\\
    \hline
    & Synthetic Scenarios & \multicolumn{5}{|c|}{}\\
    \hline
    8 & 10\% Grade          & 75 & 0.25 & 50 & - & - & -\\
    9 & 15\% Grade          & 75 & 0.25 & 40 & - & - & -\\
    10 & Strong Headwind    & 75 & 0.25 & 70 & - & - & -\\
    \hline
  \end{tabular}
  \caption{Scenario Definitions}\label{table:scenariodefs}
\end{table}
\begin{table}[H]
  \centering
  \begin{tabular}{|c|c|c|c|c|c|c|c|c|}
    \hline
    \# & $v$ ($ms^{-1}$) & $v_{wind}$ ($ms^{-1}$) & $P$ (W) & $Var$ (\%) & $\epsilon_{x}$ & $\epsilon_{l}$ & $\epsilon_{r}$ \\
  \hline
  1  & 9.2 & 0.0 & 194.67 & 0.13 & 0.00 & -0.01 & 0.01\\
  2  & 9.5 & 0.0 & 214.34 & 0.15 & 0.00 & -0.01 & 0.01\\
  3  & 6.3 & 0.0 & 294.27 & 0.69 & 0.00 & -0.06 & 0.06\\
  \hline
  4  & 7.7 & 0.0 & 228.43 & 4.46 & 0.07 & -0.21 & 0.40\\
  5  & 7.5 & 0.0 & 211.08 & 4.35 & 0.07 & -0.21 & 0.39\\
  6  & 8.3 & 0.0 & 286.05 & 3.94 & 0.06 & -0.20 & 0.35\\
  7  & 8.2 & 0.0 & 275.90 & 4.75 & 0.08 & -0.21 & 0.42\\
  \hline
  8  & 5.0 & 0.0 & 430.38 & 1.92 & 0.00 & -0.17 & 0.17\\
  9  & 3.0 & 0.0 & 344.60 & 5.34 & 0.00 & -0.47 & 0.47\\
  10 & 3.0 & 15.0 & 486.32 & 4.29 & 0.07 & -0.21 & 0.38\\
  \hline
\end{tabular}
\caption{Simulation Parameters}\label{table:scenarioresults}
\end{table}
\subsection{Simulations}
There was a degree of uncertainty in relation to the particular parameters 
chosen to recreate the experimental results from Favero's Paper. To strengthen the claims made
by this paper, 260'000 simulations were performed which capture all plausible
riding scenarios. Simulations with power values above 600w were excluded.
The input parameters were varied in the following fashion:
\begin{enumerate}
  \item \textbf{Mass}, from 1kg to 10kg in increments of 1kg (representing indoor trainer flywheel mass), and from 10kg to 90kg in increments of 5kg.
  \item \textbf{Cadence}, from 40rpm to 120rpm in increments of 10rpm.
  \item \textbf{Gradient}, from 0\% to 20\% in increments of 1\%.
  \item \textbf{Velocity}, $1ms^{-1}$ to $15ms^{-1}$ in increments of $1ms^{-1}$.
  \item \textbf{$\frac{1}{2}\rho CdA$ Values}, from $0.25m^{2}$ to $0.5m^{2}$ in increments of $0.05 m^{2}$.
\end{enumerate}
\subsubsection{AV Variation and Power Error}
\begin{figure}[H]
  \centering
  \includegraphics[scale=0.5]{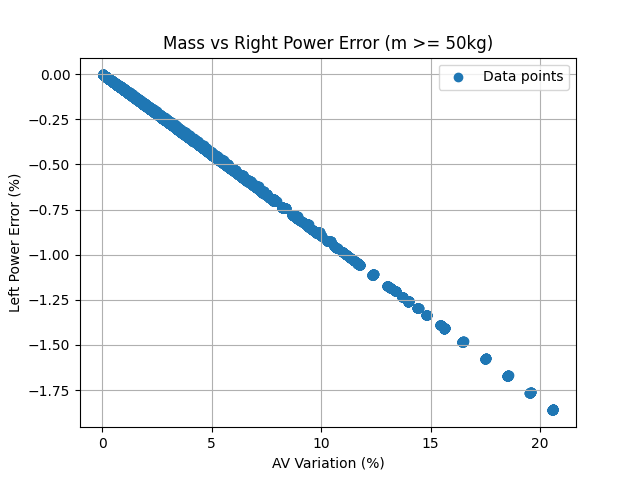}
  \includegraphics[scale=0.5]{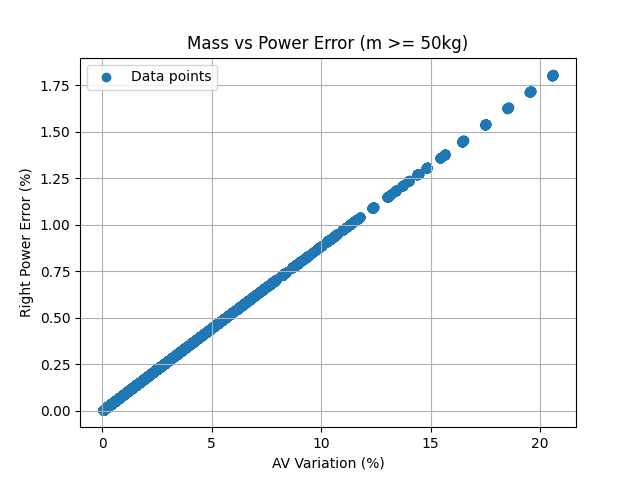}
  \caption{L/R Power Error and AV Variation}\label{fig:lrpower-av}
\end{figure}

\begin{figure}[H]
  \centering
  \includegraphics[scale=0.5]{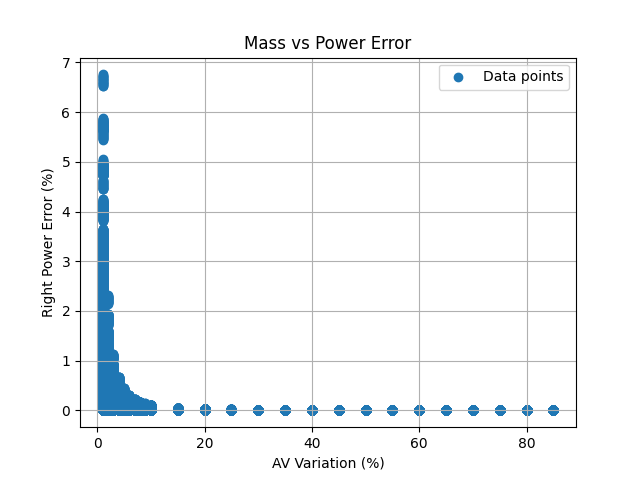}
  \includegraphics[scale=0.5]{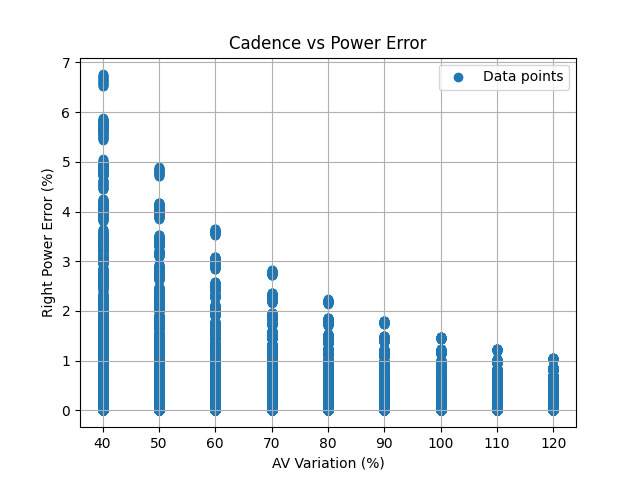}
  \caption{Relationship between Mass, Cadence and Power Error}\label{fig:masscad-powererror}
\end{figure}

Results were seperated into low and high inertia contexts. High inertia contexts
correspond to situations where a cyclist is riding outdoors. It is obvious
there there are no plausible high inertia scenarios where any degree of power error
is present, even when AV Variation is extremely high. When AV
Variation is high, indicating a very high second harmonic of AV, the phase
relationship between torque and AV is such that no real work is done by 
the interaction of the harmonics of torque and AV.
\begin{figure}[H]
  \centering
  \includegraphics[scale=0.5]{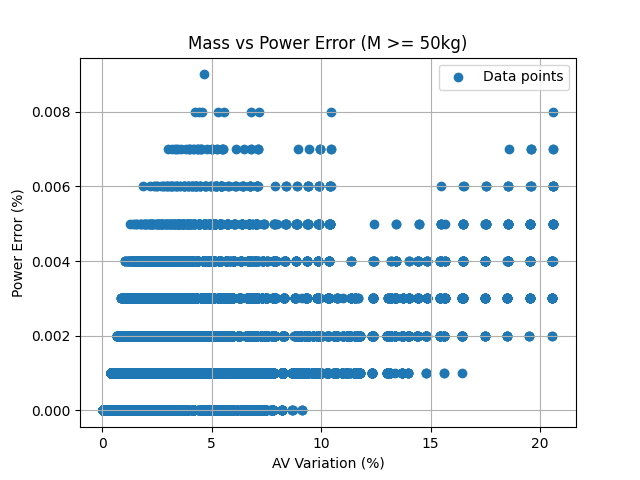}
  \includegraphics[scale=0.5]{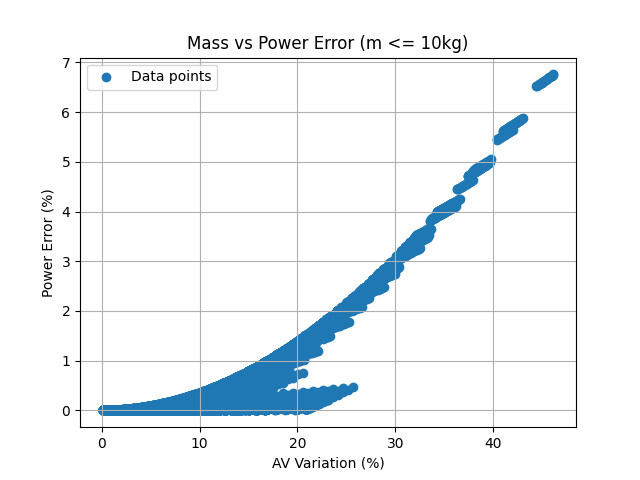}
  \caption{Power Error and AV Variation}\label{fig:powererror}
\end{figure}
Favero's Paper did in fact note high power error. Recall that power was measured unilaterally.
High levels of unilateral power error do in fact occur under simulation, as shown in Figure \ref{fig:lrpower-av},
and there is a very strong relationship between AV Variation and Unilateral Power Error.
Note the complimentarity of power error; high left power error is negated by high right power 
error in the opposing direction. The relationship between AV Variation and Unilateral Power Error
is stronger under simulation than experimentally. This could be due to measurement noise,
or rider deviation from Dynamic Equilibrium. The final consideration is that 
observed Power Error in the high inertia contexts in Table \ref{table:scenariodefs} have a positive sign; AAV overpredicts power.
This is a physical impossibility for genuine power error based on simulation results and the model
developed above. Thus, the most likely explanation for observed power error from Favero's Paper is as a 
spurious artefact of experimental design; namely, the unilateral measurement of power.
AV Variation is a poor predictor of Power Error, however Mass and Cadence 
clearly have a relationship to Power Error. As can be see seen in Figure \ref{fig:masscad-powererror}.
\section{Conclusion}
Theoretical analysis and simulation results indicate that power error resulting from the
interplay of torque and angular velocity harmonics within a pedal stroke does not 
occur to any meaningful extent in high inertia scenarios. 
Furthermore, there is no direct relationship between
variation in Angular Velocity and Power Error under conditions of Dynamic Equilibrium.
This undermines the fundamental claims
made by Favero in their paper. This paper notes the strong likelyhood of a 
flaw in Favero's experimental methodology, and strongly 
encourages Favero to perform a revised experiment accounting 
for bilateral power - this paper hypothesises that Favero will observe a substantially
lower magnitude of power error under most cirumstances where circular chainrings are used.\\
It should be noted that the claims of this paper are modest;
it is certainly possible that IAV could outperform AAV under conditions
that fall beyond the scope of this paper, which assumed Dynamic Equilibrium and the perfect
tranmission of power from the pedal to the rear wheel.
\bibliography{sources}
\section*{Annex A - Subset of Favero Tests}
\begin{table}[H]
  \centering
  \begin{tabular}{|c|c|c|c|c|c|c|c|}
    \hline
    id & Mass (kg)\tablefootnote{An estimate of 4kg was taken to represent the equivelant inertia of the Elite Quobo Magnetic Flywheel. Mass is as provided in Favero, and does not include the mass of the bicycle, clothes, etc (which is usually taken to be 10kg)}& Gradient (\%) & Cadence (rpm)\tablefootnote{Cadence was not provided for the climbing segment, so an estimate of 60rpm was taken} & AV Var (\%) & IAV (W) & AAV (W) & $\epsilon$ (\%)\\
    \hline
    1 & 4 & 0 & 90 & 4.8 & 258.5 & 257.4 & -0.42\\
    2 & 4 & 0 & 90 & 4.3 & 180.6 & 180.7 & 0.04\\
    3 & 4 & 0 & 90 & 6.1 & 261.3 & 259.4 & -0.71\\
    4 & 4 & 0 & 90 & 3.1 & 216.1 & 215.6 & - 0.23\\
    5 & 4 & 0 & 90 & 5.6 & 234.8 & 233.7 & -0.49\\
    \hline
    6 & 4 & 0 & 110 & 3.8 & 225.6 & 225.4 & -0.11\\
    7 & 4 & 0 & 110 & 4.1 & 188.7 & 189.0 & 0.16\\
    8 & 4 & 0 & 110 & 4.3 & 256.2 & 255.1 & -0.43\\
    9 & 4 & 0 & 110 & 2.6 & 172.8 & 172.7 & -0.06\\
    10 & 4 & 0 & 110 & 6.1 & 222.6 & 221.1 & -0.66\\
    \hline
    11 & 4 & 0 & 90 & 5.6 & 306.8 & 305.3 & -0.51\\
    12 & 4 & 0 & 90 & 5.6 & 264.1 & 263.8 & -0.09\\
    13 & 4 & 0 & 90 & 7.3 & 326.3 & 322.6 & -1.15\\
    14 & 4 & 0 & 90 & 3.9 & 276.6 & 276.0 & -0.21\\
    15 & 4 & 0 & 90 & 6.2 & 262.7 & 261.2 & -0.57\\
    \hline
    16 & 4 & 0 & 70 & 7.4 & 286.0 & 283.8 & -0.76\\
    17 & 4 & 0 & 70 & 10.0 & 234.1 & 232.1 & -0.82\\
    18 & 4 & 0 & 70 & 10.3 & 335.4 & 330.0 & -1.59\\
    19 & 4 & 0 & 70 & 6.5 & 261.1 & 259.9 & -0.47\\
    20 & 4 & 0 & 70 & 7.6 & 234.1 & 232.6 & -0.64\\
    \hline
    21 & 65 & 0 & 100 & 5.0 & 205.8 & 205.9 & 0.07\\
    22 & 71 & 0 & 100 & 6.0 & 161.3 & 161.2 & -0.10\\
    23 & 68 & 0 & 100 & 5.4 & 245.7 & 246.0 & 0.11\\
    24 & 67 & 0 & 100 & 4.4 & 168.8 & 169.0 & 0.15\\
    25 & 62 & 0 & 100 & 10.1 & 178.8 & 178.8 & -0.04\\
    \hline
    26 & 65 & 5 & 60 & 7.8 & 331.4 & 332.5 & 0.34\\
    27 & 71 & 5 & 60 & 10.0 & 268.9 & 270.1 & 0.45\\
    28 & 68 & 5 & 60 & 8.0 & 357.2 & 259.4 & 0.62\\
    29 & 67 & 5 & 60 & 6.9 & 258.8 & 260.4 & 0.63\\
    30 & 62 & 5 & 60 & 7.9 & 259.5 & 260.3 & 0.31\\
    \hline
    31 & 65 & 5 & 90 & 5.9 & 241.6 & 241.9 & 0.11\\
    32 & 71 & 5 & 90 & 6.5 & 215.1 & 215.1 & -0.02\\
    33 & 68 & 5 & 90 & 5.0 & 262.1 & 262.6 & 0.20\\
    34 & 67 & 5 & 90 & 4.7 & 165.8 & 165.9 & 0.06\\
    35 & 62 & 5 & 90 & 11.6 & 179.9 & 180.1 & 0.13\\
    \hline
  \end{tabular}
  \caption{Subset of Favero Tests}\label{table:subset}
\end{table}
\begin{table}[H]
  \centering
  \begin{tabular}{lllllllllll}
    \hline
     & Mass & Grade & RPM & Var & IAV & AAV & Force & FMR & E & E2 \\
    \hline
    Mass & - & 0.00 & 0.23 & 0.11 & 0.35 & 0.18 & 0.64 & 0.00 & 0.00 & 0.02 \\
    Grade & 0.00 & - & 0.00 & 0.05 & 0.43 & 0.68 & 0.02 & 0.00 & 0.00 & 0.30 \\
    RPM & 0.23 & 0.00 & - & 0.00 & 0.00 & 0.00 & 0.00 & 0.68 & 0.60 & 0.01 \\
    Var & 0.11 & 0.05 & 0.00 & - & 0.09 & 0.13 & 0.00 & 0.89 & 0.46 & 0.01 \\
    IAV & 0.35 & 0.43 & 0.00 & 0.09 & - & 0.00 & 0.00 & 0.02 & 0.14 & 0.00 \\
    AAV & 0.18 & 0.68 & 0.00 & 0.13 & 0.00 & - & 0.00 & 0.00 & 0.04 & 0.00 \\
    Force & 0.64 & 0.02 & 0.00 & 0.00 & 0.00 & 0.00 & - & 0.29 & 0.98 & 0.00 \\
    FMR & 0.00 & 0.00 & 0.68 & 0.89 & 0.02 & 0.00 & 0.29 & - & 0.00 & 0.00 \\
    E & 0.00 & 0.00 & 0.60 & 0.46 & 0.14 & 0.04 & 0.98 & 0.00 & - & 0.00 \\
    E2 & 0.02 & 0.30 & 0.01 & 0.01 & 0.00 & 0.00 & 0.00 & 0.00 & 0.00 & - \\
    \hline
    \end{tabular}    
\caption{Subset P-Values}\label{table:pvalues}
  \end{table}
\end{document}